\documentclass[aps,prl,floatfix,twocolumn,tightenlines,nobibnotes,amsmath,amssymb]{revtex4}

\usepackage{graphicx}
\usepackage{amsmath}
\usepackage{amssymb}
\usepackage{epstopdf}
\usepackage{color}
\usepackage{natbib}

\newcommand{\bra}[1] {\langle #1|}
\newcommand{\ket}[1] {|#1 \rangle}

\newcommand*{\rom}[1]{\expandafter\@slowromancap\romannumeral #1@}

\begin{document}

\title{Singlet-triplet minus mixing and relaxation lifetimes in a double donor dot}
\author{S. K. Gorman}
\affiliation{Centre of Excellence for Quantum Computation and Communication Technology, School of Physics, University of New South Wales, Sydney, New South Wales 2052, Australia}
\author{M. A. Broome}
\affiliation{Centre of Excellence for Quantum Computation and Communication Technology, School of Physics, University of New South Wales, Sydney, New South Wales 2052, Australia}
\author{M. G. House}
\affiliation{Centre of Excellence for Quantum Computation and Communication Technology, School of Physics, University of New South Wales, Sydney, New South Wales 2052, Australia}
\author{S. J. Hile}
\affiliation{Centre of Excellence for Quantum Computation and Communication Technology, School of Physics, University of New South Wales, Sydney, New South Wales 2052, Australia}
\author{\\J. G. Keizer}
\affiliation{Centre of Excellence for Quantum Computation and Communication Technology, School of Physics, University of New South Wales, Sydney, New South Wales 2052, Australia}
\author{D. Keith}
\affiliation{Centre of Excellence for Quantum Computation and Communication Technology, School of Physics, University of New South Wales, Sydney, New South Wales 2052, Australia}
\author{T. F. Watson}
\altaffiliation{Current address: Kavli Institute of NanoScience, TU Delft, P.O. Box 5046, 2600 GA Delft, The Netherlands}
\affiliation{Centre of Excellence for Quantum Computation and Communication Technology, School of Physics, University of New South Wales, Sydney, New South Wales 2052, Australia}
\author{W. J. Baker}
\affiliation{Centre of Excellence for Quantum Computation and Communication Technology, School of Physics, University of New South Wales, Sydney, New South Wales 2052, Australia}
\author{M. Y. Simmons}
\affiliation{Centre of Excellence for Quantum Computation and Communication Technology, School of Physics, University of New South Wales, Sydney, New South Wales 2052, Australia}
\date{\today}

\begin{abstract}
We measure singlet-triplet mixing in a precision fabricated double donor dot comprising of 2 and 1 phosphorus atoms separated by $16{\pm}1$~nm. We identify singlet and triplet-minus states by performing sequential independent spin readout of the two electron system and probe its dependence on magnetic field strength. The relaxation of singlet and triplet states are measured to be $12.4{\pm}1.0$~s and $22.1{\pm}1.0$~s respectively at $B_z{=}2.5$~T.
\end{abstract}

\maketitle
Single electrons confined to donor atoms in silicon are ideal candidates for semiconductor spin qubits, due to their long spin relaxation~\cite{morello2010,watson2015} and coherence lifetimes~\cite{PhysRevB.68.193207, pla2012}. In these systems the coupling of two electrons offers not only the ability to perform quantum logic between multiple qubits~\cite{nowack2011,koh2012}, but also the potential for different qubit types~\cite{hanson2007}. For example, singlet-triplet-zero qubits are appealing as they are immune to fluctuations in local magnetic field strength~\cite{petta2005} and singlet-triplet-minus qubits can be operated via electrical gates alone in the presence of a sufficiently large inhomogeneous magnetic field~\cite{PhysRevB.92.045403}.

In donor based devices the operation of multi-electron systems relies heavily on the ability to engineer a precise value of coupling between them~\cite{Wang:2016qf}. In cases where exchange coupling is desired, due to its exponential dependence on inter-donor distance, the placement of donors inside the silicon lattice must be done with close to atomic precision~\cite{PhysRevB.91.235318,PhysRevB.89.235306}. This can be achieved using scanning tunnelling microscopy (STM) hydrogen resist lithography, which has previously shown this level of donor placement~\cite{fuechsle2012}.

\begin{figure}[t!]
\begin{center}
\includegraphics[width=1\columnwidth]{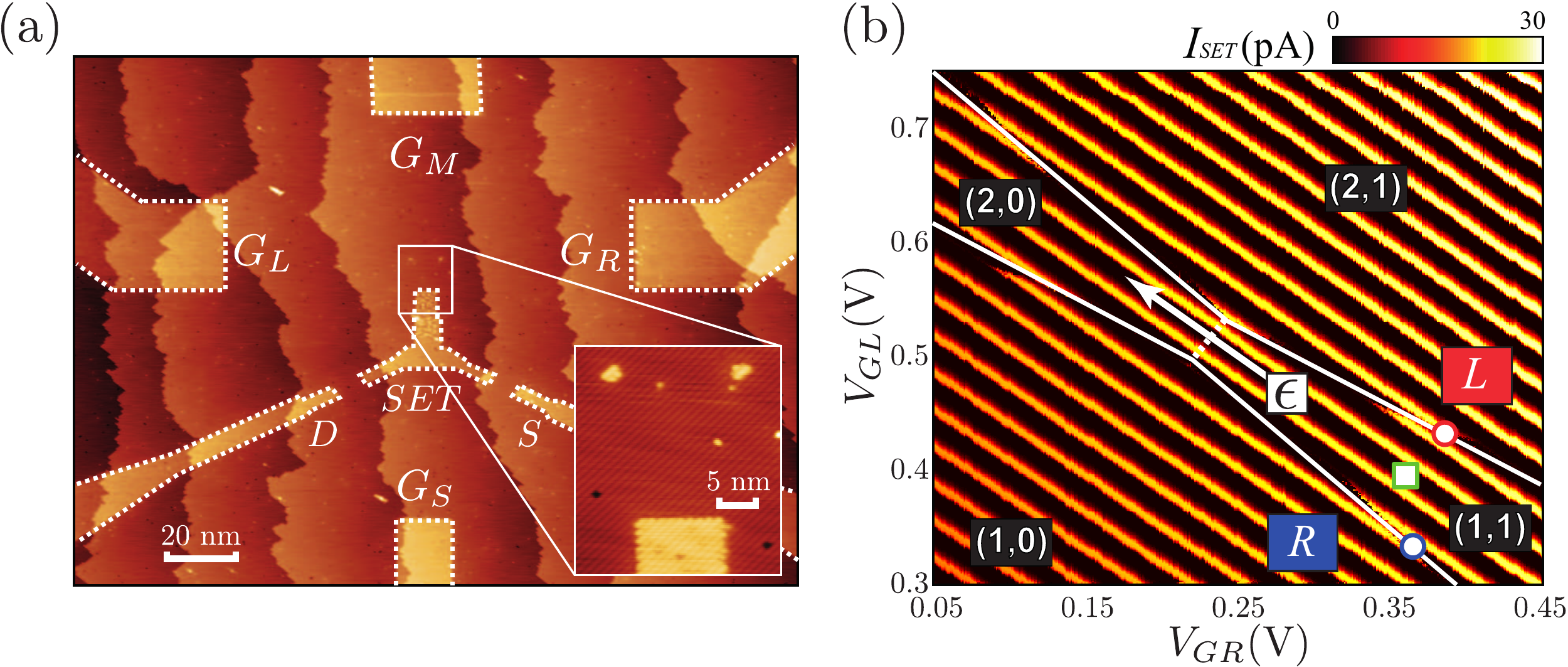}
\end{center}
\vspace{-0.5cm}
\caption{{\bf Precision fabricated donor double dot device}. a) A scanning tunnelling micrograph of the device. Two donor dots $L$ and $R$ separated by $16.0{\pm}1.0$~nm, are comprised of 2 and 1 P donors respectively. Surrounding the donor dots are three gates $G_L$, $G_M$ and $G_R$ used to control their electrostatic environment. Approximately $20$~nm from the dots is a single-electron-transistor (SET) which itself is a larger quantum dot tunnel coupled to source and drain leads. The SET also acts as an electron reservoir for $L$ and $R$ and is controlled by the gate $G_S$. Inset shows a close up of the dots and the top of the SET island. b) A charge stability map of the device shown by the transport current through the SET as a function of the gates $\{G_L,G_R\}$ near the inter-dot charge transition (1,1)-(2,0). Current peaks running at $\sim45^\circ$ show the Coulomb blockade of the SET and breaks in these correspond to transitions of single electrons from either dots $L$ or $R$, indicated by the solid white lines. The detuning axis $\epsilon$ is given by the white arrow (with a lever arm along the axis, $\alpha_{\epsilon}{=}0.071$~eV/V), and the approximate spin-readout positions for $L$ and $R$ are shown by the red and blue circles, respectively. The lever arms of the dots along the spin readout axis (same as the detuning axis) are $\alpha_{L}{=}0.041$~eV/V and $\alpha_{R}{=}0.030$~eV/V. Only one other charge transition corresponding to dot-$L$ is seen at much lower gate voltages, leading to the assignment of the charge regions shown here.}
\label{fig:device}
\end{figure}

While the relaxation of single electrons confined to donors has been well understood for over 40 years~\cite{PhysRev.118.1523, PhysRev.118.1534, PhysRev.124.1068, PhysRev.155.816} there has been far less attention to two-electron spin relaxation processes~\cite{borhani2010,PhysRevB.96.115444} since most two-donor papers focus on calculating the exchange energy, $J$~\cite{koiller2002,wellard2003}. Theoretical calculations of two-electron relaxation in Ref.~\cite{borhani2010} predict a $J^{3}$ dependence on the spin relaxation rate, whereas the inclusion of spin-orbit coupling predicts a $J^{4}$ or $J^{5}$ depending on the temperature~\cite{PhysRevB.96.115444}. Results from gate-defined quantum dots also predict a dependence of the singlet-triplet relaxation~\cite{taylor2007} on the exchange (or detuning)~\cite{johnson2005,PhysRevB.72.161301} and magnetic field~\cite{johnson2005,PhysRevLett.108.046808}. Previous experimental studies have measured singlet-triplet relaxation rates of strongly coupled electrons confined to donors, Ref.~\cite{dehollain2014} (exchange energy of $J{\approx}300\mu$eV) and Ref.~\cite{House:2015rz} (tunnel coupling $t_c{=}47\mu$eV), finding triplet-state relaxation rates of the order 100~ms and 60~ns respectively. In this later work, an STM fabricated double donor dot was probed in transport~\cite{House:2015rz} where the interaction of the source and drain leads with the double dot itself led to the observed fast spin relaxation of the singlet-triplet states. 

In this paper we use a charge sensed double donor dot fabricated with precision placement by means of STM lithography to probe singlet-triplet relaxation of a two-electron state at the (1,1)-(2,0) inter-dot charge transition. The two dots $L$ and $R$ comprise of 2 and 1 P donors respectively. The details of device fabrication and characterisation have been reported previously in Ref.~\cite{PhysRevLett.119.046802}. Figure~\ref{fig:device}a shows a close up STM image of the device which comprises of the two donor dots both of which are tunnel coupled to a larger quantum dot of approximately 1000P atoms. This larger dot is itself tunnel coupled to source ($S$) and drain ($D$) leads, across which we apply a bias of $V_{SD}{=}2.5$mV such that it operates as a single electron transistor charge sensor. Four phosphorus doped gates $\{G_L,G_M,G_R,G_S\}$ surround the device and are used for electrical control. Figure~\ref{fig:device}b shows Coulomb blockade through the charge sensing SET is shown as a function of gate volgates $V_{GL}$ and $V_{GR}$. Breaks in the current peaks represent charge transitions from either dot-$L$ or -$R$, which are used for spin-readout of these dots at the positions marked by the red and blue circles in Fig.~\ref{fig:device}b respectively. The device was measured in a dilution refrigerator with a base temperature of 100 mK and electron temperature ${\sim}$200 mK~\cite{morello2010}.

We perform single-shot spin readout of the single electrons confined to dots-$L$ and -$R$ at the $1\rightarrow2$ and $0\rightarrow1$ transitions, the so-called $D^-$ and $D^0$ readout schemes respectively~\cite{watson2015} (see Fig.~\ref{fig:device}b). Both readout schemes rely on an energy selective tunnelling to or from the donor dot to provide a readout signal from the charge sensing SET. The $D^0$ method (employed for dot-$R$) relies on the Zeeman split spin-up electron tunnelling \textit{to} the SET following which a spin-down electron tunnels back onto the dot. Conversely, the $D^-$ scheme (for dot-$L$) relies on the fact that the chemical potentials of spin-up and -down electrons transitioning to the two-electron singlet state are also split by the Zeeman energy, allowing for an equivalent readout signal~\cite{watson2015}. The advantage of utilising these two readout schemes in conjunction is that we do not need to pulse to the equivalent  $0\rightarrow1$ transition for dot-$L$ which is over 200mV away from the (1,1)-(2,0) inter-dot charge transition.

\begin{figure}[h!]
\begin{center}
\includegraphics[width=1\columnwidth]{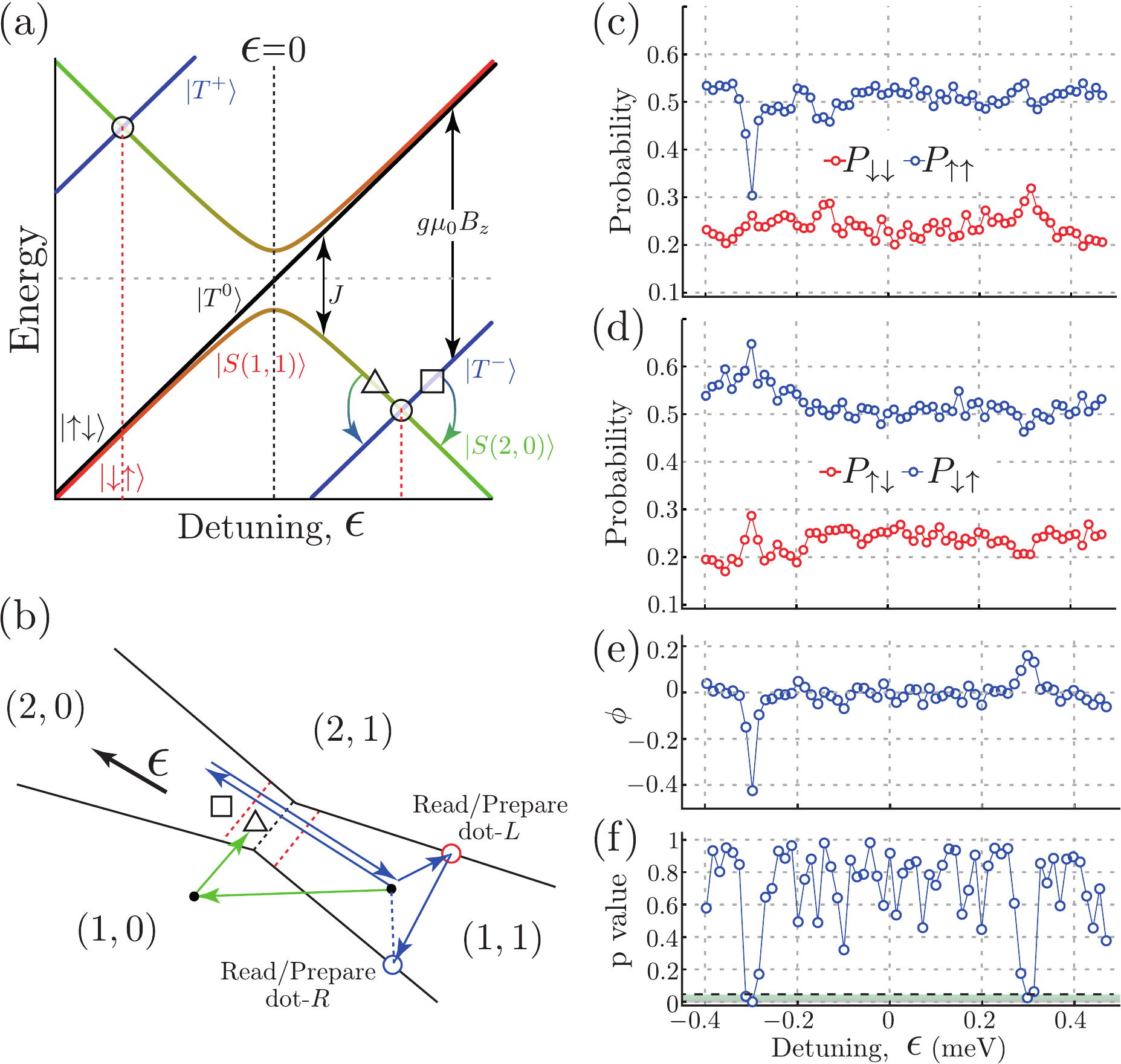}
\end{center}
\vspace{-0.5cm}
\caption{{\bf Observation of mixing between of $\ket{{S(2,0}}$ and $\{\ket{T^+},\ket{T^-}\}$ states}. (a), The eigenspectrum of the two-electron Hamiltonian showing the dependence of the singlet-triplet states as a function of detuning. Mixing between the $\{\ket{{T^+}},\ket{{T^-}}\}{\leftrightarrow}\ket{{S(2,0)}}$ occur at the positions marked by white circles. (b), The pulse sequence used to perform the two different relaxation measurements. The green line shows the pulse to load the $\ket{S(2,0)}$ state via the (1,0) charge region and measure the relaxation of the $\ket{S(2,0)}{\rightarrow}\ket{T^-}$ at the triangle. The blue line shows the pulse to measure the relaxation of $\ket{T^-}{\rightarrow}\ket{S(2,0)}$ at the square position in detuning. (c) and (d), To further investigate the mixing between the singlet and triplet states, we prepare the fully mixed state $\rho_I$ and apply a 50~ms pulse along $\epsilon$ after which we measure the probabilities (c): $P_{\downarrow\downarrow}$ and $P_{\uparrow\uparrow}$ (offset by $+0.25$ for clarity) as well as (d): $P_{\uparrow\downarrow}$ and $P_{\downarrow\uparrow}$ (offset by +0.25). (e), The correlation coefficient $\phi$ as a function of detuning for the data in (c) and (d). For $\epsilon{\sim}-0.3$~meV a decrease in $\phi$ indicates the mixing between $\ket{{T^+}}{\leftrightarrow}\ket{S(2,0)}$, whereas an increase in $\phi$ at $\epsilon{\sim}0.3$~meV is where $\ket{{T^-}}{\leftrightarrow}\ket{S(2,0)}$ occurs. (f), The p-value of a $\chi^2$ test as a function of $\epsilon$ showing statistically significant mixing of the electron spins, $p{\ll}0.05$ (green shaded region) at $\epsilon{=}{\pm} 0.3$ meV.}
\label{fig:ST_anti}
\end{figure}

The eigenspectrum of two exchange coupled electrons in a magnetic field is shown in Fig.~\ref{fig:ST_anti}a. The spin-1 triplet-states $\ket{{T^+}}$ and $\ket{{T^-}}$ are split from the $\ket{{T^0}}$ by the Zeeman $g\mu_0B_z$, crossing the singlet states at seperate detuning points where mixing between them can occur. This process necessarily involves a spin-flip between one electron spin and a surrounding nuclear spin driven by the hyperfine interaction. Since the P nuclear spins at sites $L$ and $R$ are limited in our case to $\le2$ each, an equal number of electron-nuclear spin flips can occur within the $T_1$ times of the P nuclei ($\sim$minutes for natural silicon~\cite{pla2013}). Therefore, the process driving $\ket{T^-}{\leftrightarrow}\ket{S}$ mixing will most likely be dominated by a constantly fluctuating $^{29}$Si Overhauser field~\cite{assali2011} near the $\ket{T^-}{\leftrightarrow}\ket{S}$ anticrossing. It is been recently predicted that the spin-orbit coupling could play a significant role in the two-electron relaxation processes of donors for large exchange~\cite{PhysRevB.96.115444}; however, for the tunnel coupling in this device, which was estimated to be ${\sim}200$ MHz the hyperfine driven relaxation should still dominate the relaxation process~\cite{Broome2018}.

To see this mixing process we start by preparing a random mixture of spin-up and -down on both $L$ and $R$, such that we have equal populations of all four two-electron product states,
\begin{equation}
\rho_I = \frac{\ket{{\uparrow\uparrow}}\bra{{\uparrow\uparrow}}+\ket{{\uparrow\downarrow}}\bra{{\uparrow\downarrow}}+\ket{{\downarrow\uparrow}}\bra{{\downarrow\uparrow}}+\ket{{\downarrow\downarrow}}\bra{{\downarrow\downarrow}}}{4}.
\label{eq:mixed}
\end{equation}
A 50~ms pulse is applied from the mid point between the two readout positions along the detuning axis $\epsilon$ towards the (1,1)-(2,0) charge degeneracy. Following this we return to the readout positions and perform single-shot spin-readout at $L$ and $R$ in that order, a schematic representation of the full pulsing sequence is shown in Fig.~\ref{fig:ST_anti}b. The sequence is repeated 100 times for different values of $\epsilon$ and the joint spin-readout probabilities $P_{ij}$ are shown in Fig.~\ref{fig:ST_anti}c and d. The probability of $P_{\uparrow\uparrow}$ in Fig.~\ref{fig:ST_anti}c shows a sharp decrease at approximately $-0.3$~meV, whereas both $P_{\uparrow\downarrow}$ and $P_{\downarrow\uparrow}$ show an increase at this value of detuning indicating mixing between $\ket{{T^+}}{\leftrightarrow}\ket{S(2,0)}$ (see Fig.~\ref{fig:ST_anti}a). The opposite effect is observed for mixing between $\ket{{T^-}}{\leftrightarrow}\ket{S(2,0)}$ at approximately $0.3$~meV. The position of the two singlet-triplet mixing points is used to calibrate the zero detuning point $\epsilon{=}0$.

These results can be understood by considering the relaxation processes occurring in the system. Any increase in $\ket{S(2,0)}$ due to mixing between $\ket{{T^+}}{\leftrightarrow}\ket{S(2,0)}$ at negative detuning (white circle at $\epsilon{<}0$ in Fig.~\ref{fig:ST_anti}a) will result in a fast charge relaxation from $\ket{S(2,0)}{\rightarrow}\ket{S(1,1)}$ leading to a \textit{decrease} in the overall $\ket{{T^+}}$ population after $\ket{{T^+}}{\leftrightarrow}\ket{S(2,0)}$ mixing. However, at positive detuning, relaxation from the $\ket{S(1,1)}{\rightarrow}\ket{S(2,0)}$ (as well as $\ket{{T^0}}{\rightarrow}\ket{S(2,0)}$), will result in an \textit{increase} in $\ket{{T^-}}$ due to mixing between $\ket{{T^-}}{\leftrightarrow}\ket{S(2,0)}$ (white circle at $\epsilon{>}0$ in Fig.~\ref{fig:ST_anti}a). The resulting correlation coefficient $\phi$, as well as its statistical significance, are given in Fig.~\ref{fig:ST_anti}e and f respectively. A significantly negative value of $\phi$ for $\epsilon{<}0$ and a positive value for $\epsilon{>}0$ follows as a result of the mixing described above, furthermore, its values at these positions are statistically significant with a $\chi^2$ p-value of $\ll0.05$ in both cases.
\begin{figure}[t!]
\begin{center}
\includegraphics[width=1\columnwidth]{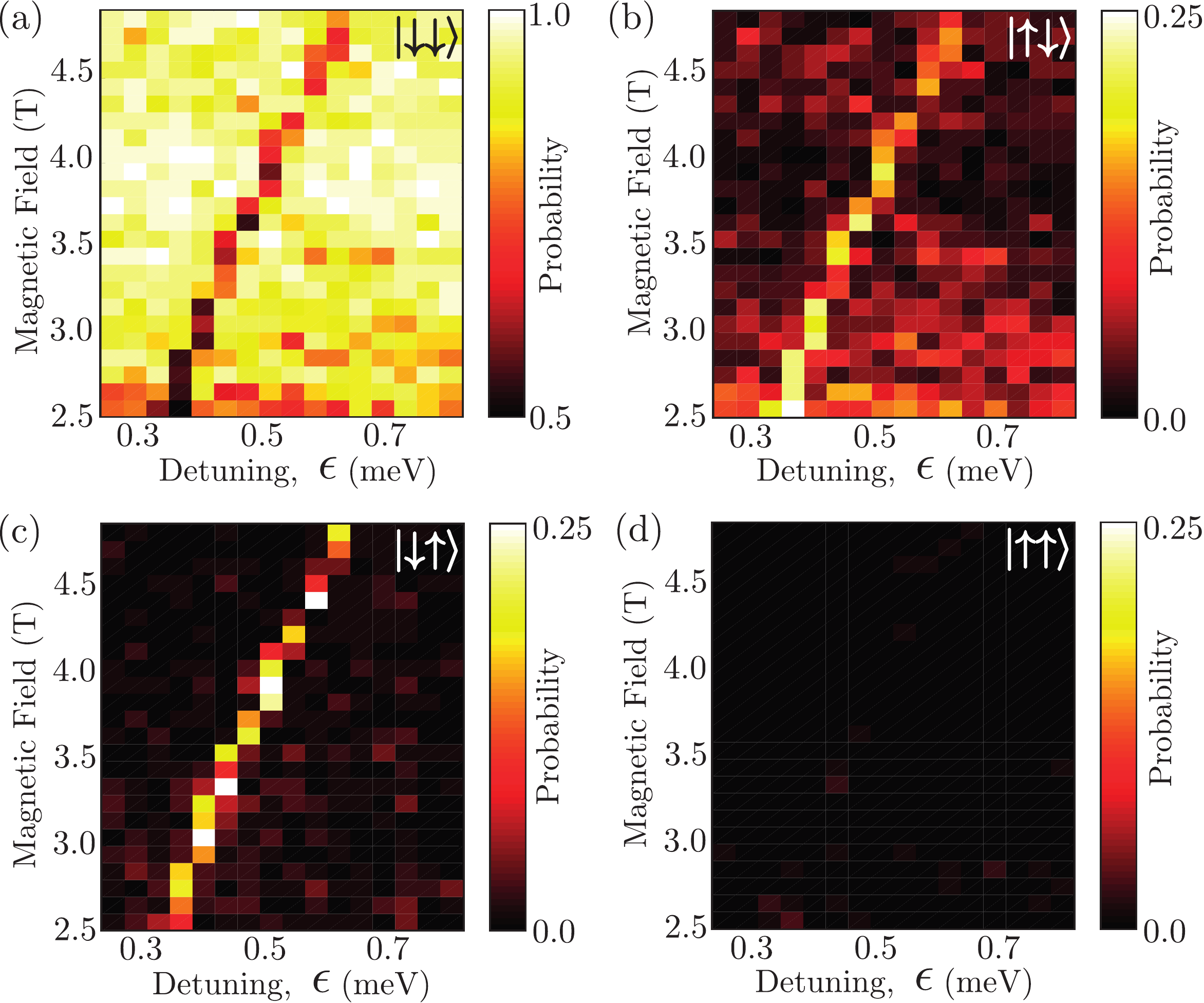}
\end{center}
\vspace{-0.5cm}
\caption{{\bf Mixing between $\ket{{S(2,0)}}$ and $\ket{{T^-}}$ as a function of magnetic field and detuning}. The initial state $\ket{{T^-}}{=}\ket{{\downarrow\downarrow}}$ is prepared by means of reading both the left and right dot electrons, after which a $50$~ms pulse is applied along the detuning axis $\epsilon$. Upon returning to the readout positions, the probabilities $\{P_{\downarrow\downarrow},P_{\uparrow\downarrow},P_{\downarrow\uparrow},P_{\uparrow\uparrow}\}$ are measured. Mixing between the $\ket{{T^-}}{\leftrightarrow}\ket{{S(2,0)}}$ is apparent where a decrease in $P_{\downarrow\downarrow}$ is observed shown in (a), at the expense of an increase in the anti-correlated probabilities $P_{\uparrow\downarrow}$ and $P_{\downarrow\uparrow}$ shown in (b) and (c), respectively. (d), The probability $P_{\uparrow\uparrow}$ is not affected and therefore shows no significant change. The magnitude of the exchange energy, $J{=}\gamma_\textrm{e}B_z$ can be seen to depend linearly on the corresponding detuning value, $\epsilon$.}
\label{fig:Bz_dep}
\end{figure}

Providing further evidence that the above results are consistent with the $\ket{T^-}{\leftrightarrow}\ket{S(2,0)}$ mixing, we repeat the equivalent pulsing scheme with the initial state $\rho_i{=}\ket{{\downarrow\downarrow}}\bra{{\downarrow\downarrow}}$ for different values of static magnetic field, $B_z$. This state can be prepared deterministically, as the spin readout process leaves a spin-down electron. The results of this experiment are shown in Fig.~\ref{fig:Bz_dep}. We see that from Fig.~\ref{fig:ST_anti}a that the position of singlet-triplet minus mixing occurs where the exchange energy is equal to the Zeeman splitting, that is, where $J{=}g\mu_0B_z$, which can clearly be seen in our data. Importantly, for devices where the tunnel coupling $t_c$ is of the order of the Zeeman, a non-linear dependence of singlet-triplet mixing should be observed as in the so called \textit{spin-funnel} experiments~\cite{maune2012}. However, the measurements shown in Fig.~\ref{fig:Bz_dep} are restricted to $B_z{\ge}2.5$~T in order that high fidelity spin readout can be performed, the position of $\ket{T^-}{\leftrightarrow}\ket{S(2,0)}$ mixing therefore shows a linear dependence on $B_z$, placing a bound on the tunnel coupling is $t_c\ll2.5g\mu_0$.

\begin{figure}[h!]
\begin{center}
\includegraphics[width=0.75\columnwidth]{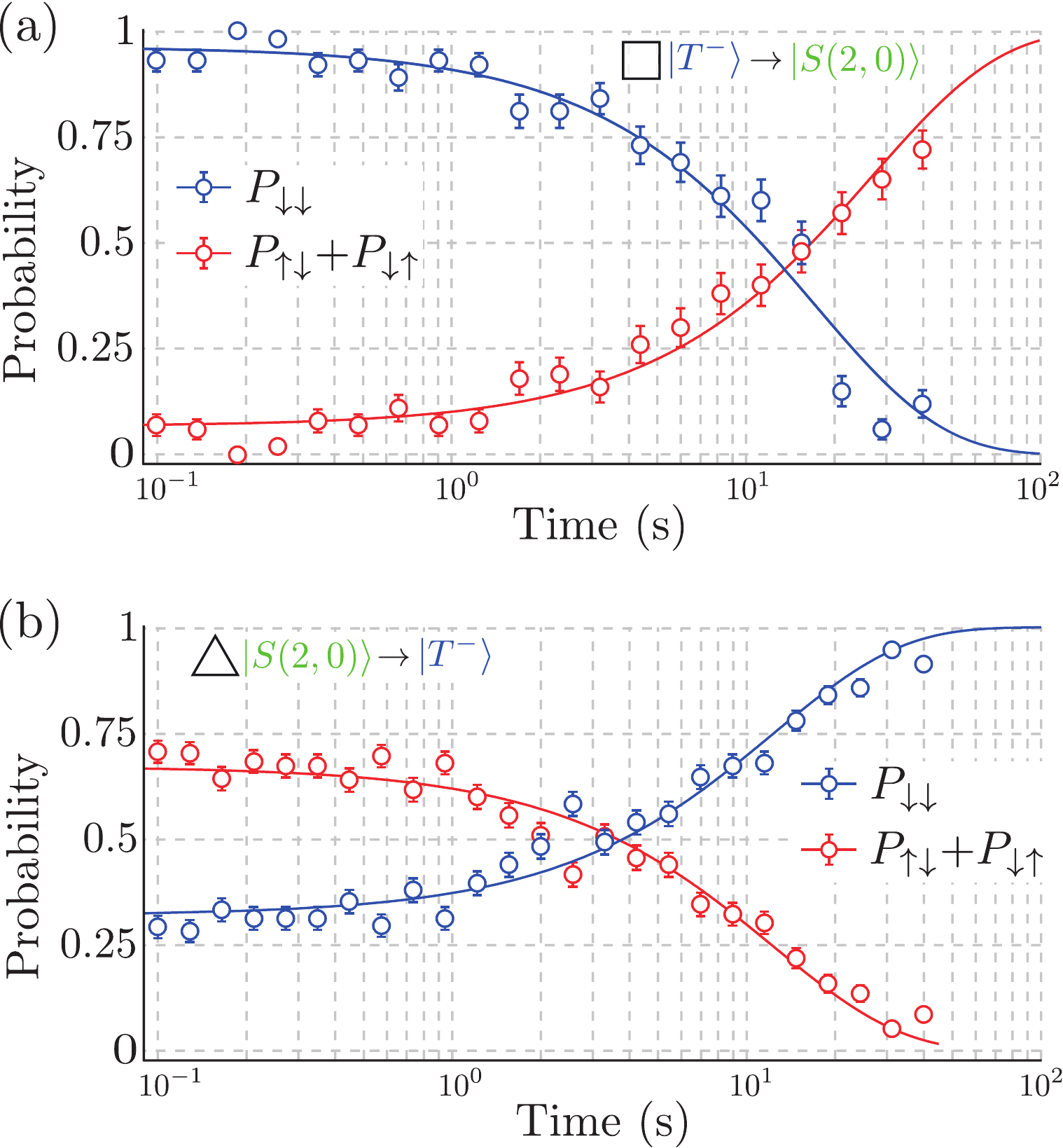}
\end{center}
\vspace{-0.5cm}
\caption{{\bf Relaxation of singlet and triplet states as measured by two-electron spin readout}. (a), The relaxation of $\ket{T^-}$ into $\ket{S(2,0)}$ measured at $B_z{=}2.5$~T. The probability of measuring anti-correlated electron spins $P_{\uparrow\downarrow}$ and $P_{\downarrow\uparrow}$ increases as a function of wait time indicating relaxation into the singlet state. A fit to the data gives a relaxation time, $T_1{=}22.1{\pm}1.0$ s. (b), Relaxation of $\ket{S(2,0)}$ into $\ket{T^-}$ can be seen by an increasing $P_{\downarrow\downarrow}$ probability with wait time. A fit to the data gives a relaxation time, $T_1{=}12.4{\pm}1.0$ s.}
\label{fig:ST_relax}
\end{figure}

In this last section we present results of singlet-triplet relaxation. The relaxation of singlet-triplet states can both be probed around the $\ket{T^-}{\leftrightarrow}\ket{S(2,0)}$ mixing point, see Fig.~\ref{fig:ST_relax}a. After deterministically loading $\ket{T^-}$ we pulse to a position 1mV positively detuned from the $\ket{T^-}{\leftrightarrow}\ket{S(2,0)}$ mixing point where we wait up to 40~s, see Fig.~\ref{fig:ST_relax}a. Here the $\ket{S(2,0)}$ is the ground state and we see that the probabilities $P_{\uparrow\downarrow}$ and $P_{\downarrow\uparrow}$ increase as $P_{\downarrow\downarrow}$ decreases as a result of relaxation with a $T_1{=}22.1{\pm}1.0$~s. To deterministically prepare the singlet state, $\ket{S(2,0)}$, we perform the pulse sequence shown by the green arrows in Fig.~\ref{fig:ST_anti}b. After unloading an electron from $R$ by pulsing into the (1,0) charge region we pulse to the (2,0) charge region 1mV \textit{negatively} detuned from the $\ket{T^-}{\leftrightarrow}\ket{S(2,0)}$ mixing point. In the (2,0) region only a singlet state can be loaded due to Pauli spin blockade~\cite{House:2015rz,weber2014}. At this position the $\ket{T^-}$ is the ground state, and relaxation of the singlet state occurs with $T_1{=}12.4{\pm}1.0$~s, see Fig.~\ref{fig:ST_relax}b. Note that for this experimental run the spin readout fidelity of the left-dot was significantly reduced due to misalignment of the spin readout voltage level, this accounts for the increased value of $P_{\downarrow\downarrow}$ from $t{=}0$, that is, we were more likely to count $\ket{{\uparrow}}$ as $\ket{{\downarrow}}$. The value of $T_1$ measured here is in stark contrast to a previously measured relaxation time of 60~ns~\cite{House:2015rz}, where coupling to in plane leads meant the two-electron state was poorly isolated causing rapid relaxation of the singlet state via spin exchange with the leads.

The coupling of multiple electron spins is a prerequisite for scalable solid-state quantum computation. The resulting entangled states can be utilised directly by performing single-shot spin readout~\cite{nowack2011} or even as a qubit subspace~\cite{maune2012,PhysRevB.92.045403}. The long $T_1$ lifetimes of singlet and triplet states measured here do not present as a challenge for future experiments, and were expected given previous results on single spin states of donors~\cite{morello2010,watson2015,pla2012}. However, the coherent characteristics of such states are unknown and investigations of them will require either the direct manipulation in a singlet-triplet qubit subspace~\cite{PhysRevB.92.045403}, or similar protocols for the individual electron spins~\cite{nowack2011}.

\section{acknowledgements}
This research was conducted by the Australian Research Council Centre of Excellence for Quantum Computation and Communication Technology (project no. CE110001027) and the US National Security Agency and US Army Research Office (contract no. W911NF-08-1-0527). M.Y.S. acknowledges an ARC Laureate Fellowship.


\end{document}